%% file: B_meson_J_P.tex
\documentclass[10pt,sort&compress,a4paper]{article}

\usepackage{amssymb}
\usepackage{mathtools}
\usepackage{gensymb}
\usepackage{graphics}
\usepackage{url}
\usepackage{enumitem}
\usepackage{standalone}
\usepackage{dcolumn}
\usepackage[square,sort,comma,numbers]{natbib}
\usepackage{tikz}
\usetikzlibrary{decorations.pathmorphing}
\usetikzlibrary {arrows.meta}
\usetikzlibrary{calc}
\usepackage{caption}

\let\oldtabular\tabular 
\renewcommand{\tabular}{\small\oldtabular}

\begin{document}
\title{Intrinsic Angular Momentum of the B Mesons: Proposal for a Model-Independent Measurement}
\author{M.P. Fewell \cr ARC Centre of Excellence for Dark-Matter Particle Physics and \cr Department of Physics, The University of Adelaide, \cr Adelaide SA 5005, Australia}
\date{May 2023}
\maketitle

\begin{abstract}
\noindent
The intrinsic angular momentum (``spin'') of the B mesons has not so far been measured, notwithstanding the large amount of experimental attention that these particles have received in the four decades since the first of them was discovered. This paper draws attention to the applicability of a long-standing method for spin measurement from nuclear spectroscopy, concluding that known decay data point to the in-principle feasibility of a model-independent measurement of the spin of the B$^+$ meson. For the B$^0$ and B$^0_\mathrm{s}$ mesons, known decay chains allow the Standard-Model prediction to be tested, but are insufficient for a fully model-independent spin measurement. Suitable decay chains are not yet known for the B$^+_\mathrm{c}$ meson. The Standard Model predicts all B mesons to have spin zero. If this turns out not to be so, then the possibility of parity measurement opens up, using a related method.


\end{abstract}

\section{Introduction} \label{sec:Introduction}
\noindent
The first of the B mesons, comprising a b or $\mathrm{\bar{b}}$ quark paired with a lighter quark, was discovered in the early 1980s \citep{Be83}. All four flavours are now known and have been extensively studied: the available information on their masses, decay and other properties occupies hundreds of pages in the latest compilation from the Particle Data Group \citep{Wo22}. It is striking therefore that the listing for each flavour begins with the comment ``Quantum numbers not measured. Values shown are quark-model predictions'' (or equivalent) \citep{Wo22}. The present paper draws attention to the in-principle applicability of a decades-old method for measuring one of the quantum numbers, the intrinsic angular momentum (``spin'') $J$. If technically feasible, the result would be model-independent, in the sense of not relying on predictions of the Standard Model (SM) of particle physics.

\section{Conceptual Overview} \label{sec:Overview}
\noindent
The measurement of excited-state spin has been a standard task in nuclear spectroscopy since mid last century (e.g. \citep{Fr53, Kl53, Ha66, Ro67, Wa67, Kr71, Ta71, En78, Br80, Ma87, Sm19, Mc22}). The technique exploits observable effects arising from the angular-momentum conserving properties of radiative decay. This section gives an introductory overview of the method; numerical details relevant to the B mesons follow in Sect. \ref{sec:Details}.

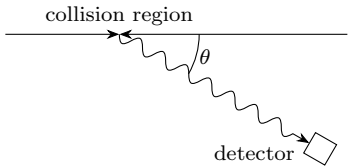
\begin{figure}[b]
    \centering
    \input{fig_angdist}
    \caption{Gamma-ray angular distribution following a reaction in colliding-beam geometry: the intensity of the radiation is measured as a function of the angle $\theta$ to the beam direction}
    \label{fig:angdist}
\end{figure}

Experimentally the simplest way of determining the angular momentum of a quantum state decaying radiatively is to measure the angular distribution of the radiation relative to some physically meaningful direction. For example, Fig. \ref{fig:angdist} sketches the application to a state formed in a particle-beam-induced reaction, where the reference direction is provided by the direction of propagation of the contra-propagating particle beams.

\begin{figure}
    \centering
    \input{fig_angcorr.tex}
    \caption{Geometry for measuring gamma--gamma angular correlations: in a cascade of two decays, the intensity of the second gamma ray in the cascade is measured relative to the propagation direction of the first using energy-resolved coincidence spectroscopy}
    \label{fig:angcorr}
\end{figure}
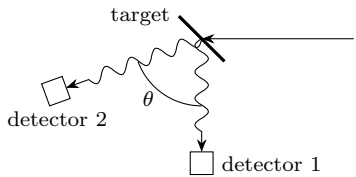

The configuration shown in Fig. \ref{fig:angdist} is inadequate for the B mesons, because these are expected to have spin zero, implying isotropic emission of the radiation, which cannot be distinguished from some other spin value $J$ that happens to have been formed with all its projection substates $M$ equally populated. The difficulty can be overcome when the state in question decays by a cascade of two radiative transitions: in general terms, the first transition produces an alignment\footnote{A state is said to be \emph{aligned} if the fractional populations $w(M)$ (with $\sum_{M} w(M) = 1$) of its projection substates $M$ are not all equal and obey $w(-M)=w(M)$. If the last relation does not hold, the state is said to be \emph{polarised}.} of the intermediate state, to an extent that depends on the $J$ values involved. The alignment results in a non-isotropic radiation pattern for the second transition. Figure \ref{fig:angcorr} shows a typical experimental setup for measuring gamma--gamma angular correlation in a fixed-target experiment with discrete gamma-ray detectors. Three of the B mesons are known to have suitable decay chains, as discussed in Sect. \ref{sec:Decays}.

The concept of the proposal in this paper follows long-established practice in nuclear spectroscopy: the experimentally observed angular correlation is compared with predictions of radiative-decay theory, which depend on the $J$ values involved, seeking a case where the observed angular correlation matches the prediction for just one set of $J$ values.

\section{Application to B-meson decay} \label{sec:Details}
\subsection{Decay Chains}
\label{sec:Decays} \noindent
Two-step radiative decay chains are known for the three lighter B mesons; Table \ref{tab:decays} collects the available data. (Information on radiative decays of the B$_\mathrm{c}$ meson is not yet available \citep{Wo22}.) Only two-body decays are considered at each step, for two reasons: it constrains the angular-momentum flow and it causes each emitted photon to have a definite energy, allowing (in principle) the photons to be distinguished from each other and from background photons by energy. With one exception---the reason for which is explained in Sect. \ref{sec:cascade}---the list is restricted to decays with measured branching-fraction values; that is, decays for which the branching fraction is known only as an upper limit are omitted.

\begin{table}[!b]
    \centering
    \caption{Candidate two-step cascade radiative decay modes for B-meson spin measurements, with branching fractions \citep{Wo22}}
    \begin{tabular}{ l @{$\to$ } l @{$\to$ } l l l }
        \noalign{\medskip}\hline\noalign{\smallskip}
        \multicolumn{3}{l}{ } & \multicolumn{2}{c}{branching fractions} \\[-1pt]
        \multicolumn{3}{c}{decay cascade} & first decay & second decay \\
        \noalign{\smallskip}\hline\noalign{\smallskip}
        B$^+$ & $\rho(770)^+\gamma$ & $\pi^+\gamma\gamma$ & $9.8 \times 10^{-7}$ & $4.5 \times 10^{-4}$ \\
        B$^+$ & K*(892)$^+\gamma$ & K$^+\gamma\gamma$ & $3.92 \times 10^{-5}$ & $9.8 \times 10^{-4}$ \\
        B$^+$ & K$^*_2(1430)^+\gamma$ & K$^+\gamma\gamma$ & $1.4 \times 10^{-5}$ & $2.4 \times 10^{-3}$ \\
        \noalign{\smallskip}\hline\noalign{\smallskip}
        B$^0$ & $\rho(770)^0\gamma$ & $\pi^0\gamma\gamma$ & $8.6 \times 10^{-7}$ & $4.7 \times 10^{-4}$ \\
        B$^0$ & $\rho(770)^0\gamma$ & $\eta\gamma\gamma$ & $8.6 \times 10^{-7}$ & $3.00 \times 10^{-4}$ \\
        B$^0$ & $\omega(782)\gamma$ & $\pi^0\gamma\gamma$ & $4.4 \times 10^{-7}$ & $0.0835$ \\
        B$^0$ & $\omega(782)\gamma$ & $\eta\gamma\gamma$ & $4.4 \times 10^{-7}$ & $4.5 \times 10^{-4}$ \\
        B$^0$ & K*(892)$^0\gamma$ & K$^0\gamma\gamma$ & $4.18 \times 10^{-5}$ & $2.46 \times 10^{-3}$ \\
        B$^0$ & K$^*_2(1430)^0\gamma$ & K$^0\gamma\gamma$ & $1.24 \times 10^{-5}$ & $< 9 \times 10^{-4}$ \\
        \noalign{\smallskip}\hline\noalign{\smallskip}
        B$^0_\mathrm{s}$ & $\phi$(1020)$\gamma$ & $\eta\gamma\gamma$ & $3.4 \times 10^{-5}$ & $0.0130$ \\
        B$^0_\mathrm{s}$ & $\phi$(1020)$\gamma$ & $\pi^0\gamma\gamma$ & $3.4 \times 10^{-5}$ & $1.32 \times 10^{-3}$ \\
        \noalign{\smallskip}\hline\noalign{\smallskip}
    \end{tabular}
    \label{tab:decays}
\end{table}

\subsection{Radiative Decay Theory with a Spin-Zero Final State}
\label{sec:Spin} \noindent
The geometries of Figs \ref{fig:angdist} and \ref{fig:angcorr} both satisfy the conditions for the variation of photon intensity with angle $\theta$ to be given by Legendre polynomials $P_k(\cos\theta)$ of even order $k$ only (e.g. \citep[Sect. III.E]{Ro67}):
\begin{equation} \label{angdist}
    W(\theta) = A_0[1 + a_2P_2(\cos\theta) + a_4P_4(\cos\theta)+\ldots],
\end{equation}
\noindent
where $a_k$ are coefficients, the theoretical expression for which is given in Eq. \eqref{ak} below for the geometry of Fig. \ref{fig:angcorr}. In general, higher-order terms can occur, but all decay cascades in Table \ref{tab:decays} have combinations of spins that require $a_6$ and all higher-$k$ coefficients to be zero; indeed for most of them $a_4$ equals zero (Sect. \ref{sec:spin1}).

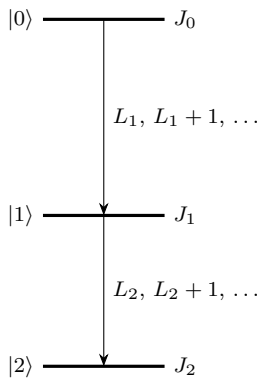
\begin{figure}[!b]
    \centering
    \input{fig_cascade.tex}
    \caption{Labelling of the angular momenta in a two-step cascade of radiative transitions: states---particles---have spin $J_i$ and photons multipolarities $L_i$}
    \label{fig:cascade}
\end{figure}

Figure \ref{fig:cascade} shows the labelling adopted here for spins and multipolarities. The multipolarities run over all values that satisfy the triangle relation with the relevant spins. Every cascade in Table \ref{tab:decays} ends in a spin-zero particle; that is, $J_2=0$. In consequence, $L_2=J_1$ is the sole multipolarity for the second transition; this simplifies the theoretical expression for $a_k$. The SM predicts that all B mesons also have spin zero, but since we aim to determine this independently of the SM, $J_0$ is not set to zero from the outset. With $J_2=0$ and $L_2$ set to $J_1$, the expression for $a_k$ is \citep[Eq. (3.73)]{Ro67}

\begin{align} \label{ak}
    a_k = &\left[R_k(L_1L_1J_1J_0)-2\delta_1R_k(L_1L_1^{\prime}J_1J_0)+\delta_1^2R_k(L_1^{\prime}L_1^{\prime}J_1J_0)\right] \nonumber \\
          & \times R_k(J_1J_1J_10)/(1+\delta_1^2),
\end{align}
\noindent
where $L_1^{\prime}=L_1+1$ and the $R_k$ coefficients are \citep[Eq. (3.36)]{Ro67}
\begin{equation} \label{Rk}
    R_k(abcd)=(-)^{c-d}\sqrt{(2a+1)(2b+1)(2c+1)}C(abk,1\,\mathord{-}1\,0)W(ccab;kd),
\end{equation}
\noindent
with $C(abc,m_1m_2m)$ being a Clebsch-Gordon coefficient in Rose's notation \citep[p. 37]{Ro57} and $W(abcd;ef)$ a Racah coefficient in standard notation. The quantity $\delta_1$ in Eq. \eqref{ak} is a \textit{mixing ratio}: the ratio of reduced transition matrix elements for the two lowest possible multipolarities of the first transition in the cascade \citep[Eq. (3.48)]{Ro67}. Its definition is such that $\delta_1=0$ means a transition with pure $L_1$ multipolarity, whereas $\delta_1=\pm \infty$ means pure $L_1+1$. The examples in the two following subsections should clarify this. In general, more than two multipolarities are possible if one of the states is spin 2, but typically a third multipolarity contributes much more weakly than the lowest two; the possibility is neglected in the following numerical analysis.

To compare experimentally measured values of $a_k$ with Eq. \eqref{ak}, one substitutes explicit values for $J_0$ and $J_1$. In the following subsections, we choose to give results for $J_0=0$, 1, 2.

\subsubsection{Spin-1 Intermediate State}
\label{sec:spin1} \noindent
All but two cascades in Table \ref{tab:decays} have $J_1=1$. For these, $a_4=0$, because a triangle relation obeyed by the Racah coefficients causes $R_4(1110)$ (second line of Eq. \ref{ak}) to be zero. Explicit expressions for $a_2$ for the three values of $J_0$ in question are
\begin{equation} \label{a21}
    a_2=\left\{
    \begin{array}{ll}
        1/2                                                                             & (J_0=0) \\[0.3em]
        \dfrac{-1+6\delta_1-\delta_1^2}{4(1+\delta_1^2)}                                & (J_0=1) \\[1em]
        \dfrac{1-6\delta_1\sqrt{5}+5\delta_1^2}{20(1+\delta_1^2)} & (J_0=2).
    \end{array}
    \right.
\end{equation}
When $J_0=0$, the first transition is pure dipole (i.e purely $L_1=1$, so that $\delta_1=0$), which is why $\delta_1$ does not appear in the first line of Eq. \eqref{a21}.

\begin{figure}[!b]
    \centering
    \includegraphics[width=0.75\textwidth]{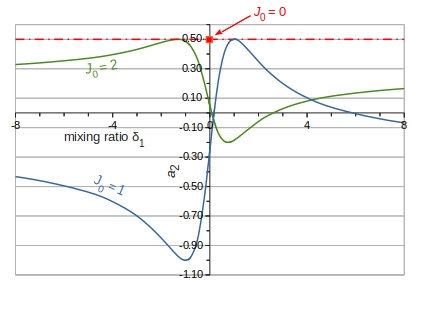}
    \captionsetup{position=below, skip=1pt}
    \caption{Dependence of $a_2$ on $\delta_1$ for decay chains with a spin-1 intermediate state (Eq. \ref{a21}). When $J_0=0$, $\delta_1$ is identically zero, so the result for that case is shown as a single point at $\delta_1=0$, with a chain line for reference.}
    \label{fig:spin1}
\end{figure}

Figure \ref{fig:spin1} shows the behaviour of Eq. \eqref{a21}. Let us suppose that $a_2$ has been measured experimentally and the expected result obtained, namely $a_2=0.5$, $a_4=0$. This would be consistent with the SM prediction of $J_0=0$, but Fig. \ref{fig:spin1} shows that it would equally be consistent with $J_0=1$, $\delta_1=1$, and with $J_0=2$, $\delta_1=-3/\sqrt{5}=-1.3416$. There may be higher $J_0$ values that also give $a_2=0.5$ for particular $\delta_1$ values. In summary, a result of $a_2=0.5$, $a_4=0$ is insufficient for a model-independent determination of $J_0$. The next subsection shows how this difficulty can be overcome if a decay chain involving a $J_1$ value different from 1 is available. Table \ref{tab:decays} includes two cascades with $J_1=2$.

\subsubsection{Spin-2 Intermediate State}
\label{sec:spin2} 
\begin{figure}[!b]
    \centering
    \includegraphics[width=0.75\textwidth]{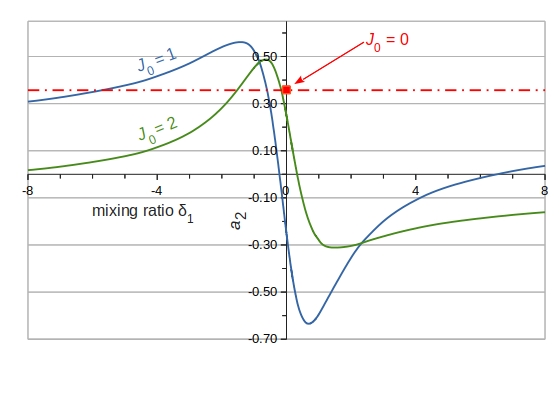}
    \captionsetup{position=below, skip=-1pt}
    \caption{Like Fig. \ref{fig:spin1}, but for decay chains with a spin-2 intermediate state (Eq. \ref{a22})}
    \label{fig:spin22}
\end{figure}

\begin{figure}
    \centering
    \includegraphics[width=0.75\textwidth]{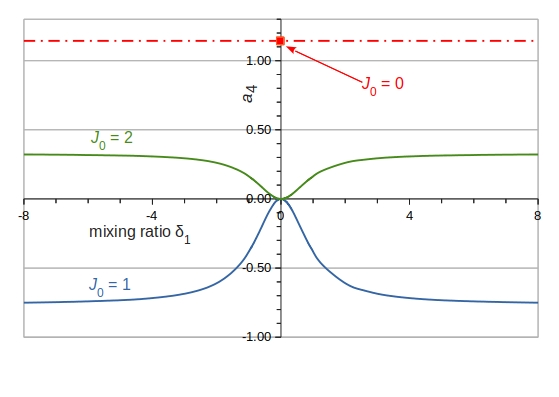}
    \captionsetup{position=below, skip=-1pt}
    \caption{Like Fig. \ref{fig:spin1}, but the dependence of $a_4$ for decay chains with a spin-2 intermediate state (Eq. \ref{a42})}
    \label{fig:spin24}
\end{figure}

\noindent
When $J_1=2$, the angular correlation gains an $a_4$ term in addition to the $a_2$ term. Explicit expressions for the three $J_0$ values under consideration are
\begin{equation} \label{a22}
    a_2=\left\{
    \begin{array}{ll}
        5/14                                                            & (J_0=0) \\[0.3em]
        \dfrac{-7-14\delta_1\sqrt{5}+5\delta_1^2}{28(1+\delta_1^2)}     & (J_0=1) \\[1em]
        \dfrac{47-14\delta_1\sqrt{105}-15\delta_1^2}{196(1+\delta_1^2)} & (J_0=2)
    \end{array}
    \right.
\end{equation}
and
\begin{equation} \label{a42}
    a_4=\left\{
    \begin{array}{ll}
        8/7                                     & (J_0=0) \\[0.3em]
        \dfrac{-16\delta_1^2}{21(1+\delta_1^2)} & (J_0=1) \\[1em]
        \dfrac{16\delta_1^2}{49(1+\delta_1^2)}  & (J_0=2).
    \end{array}
    \right.
\end{equation}
The behaviour of Eqs \eqref{a22} and \eqref{a42} is shown in Figs \ref{fig:spin22} and \ref{fig:spin24} respectively. An experimental result in agreement with the SM prediction of $J_1=0$ would be $a_2=0.357$, $a_4=1.143$. Once again the $a_2$ value is ambiguous as regards the implied $J_0$ value, but the $a_4$ value is not. Furthermore, the expected $a_4$ value for $J_0=0$ is so much larger in magnitude than any possible value for $J_0=1$ or $J_0=2$ that prospects for measurement seem promising. Detailed modelling of extant detectors could explore the extent to which the promise can be realised.

\section{Discussion}
\label{sec:Discussion} \noindent
This section addresses some practical questions in the application of the technique to the measurement of spin in the B mesons.

\subsection{The 0--2--0 Cascade}
\label{sec:cascade} \noindent
The usefulness of the behaviour of the 0--2--0 radiative cascade shown in Figs \ref{fig:spin22} and \ref{fig:spin24} has long been recognised in nuclear spectroscopy. It was exploited in precision measurements to assess the accuracy of finite-solid-angle corrections \citep{Br80} and it has for many decades been identified by the International Network for Nuclear Structure Data Evaluation as a providing a strong basis for spin assignment (cf. the identical wording of ``proposition 12'' in \citep{Ma87} from 1987 and \citep{Mc22} from 2022).

Unfortunately only the B$^+$ meson has an identified two-step radiative decay chain that meets all the criteria: a spin-2 intermediate particle, both decays two-body with the second particle a photon, and both branching fractions measured. At noted in Table \ref{tab:decays}, the B$^0$ meson has a known decay path through the corresponding K$^*_2$ meson that would be suitable, but the branching fraction for the second step is known only as an upper limit. The need for a firm value is obvious: should it turn out to be not much less than the upper limit, a model-independent measurement of the spin of the B$^0$ meson then becomes possible (in principle). A two-body decay path of the B$^0_\mathrm{s}$ meson involving a K$^*_2$ is known, but the accompanying particle is a K meson, not a photon \citep{Wo22}. It may be possible to adapt the analysis of Rose and Brink \citep{Ro67} to this case; if so the result must be expected to differ from the behaviour in Figs \ref{fig:spin22} and \ref{fig:spin24} because K mesons are spin-zero.

\subsection{Indicative Decay-Chain Production Rates}
\label{sec:Rates} \noindent
The rate at which one of the decay chains in Table \ref{tab:decays} is produced in a particle-beam experiment can be estimated as the product of
\begin{itemize}[itemsep=0pt, parsep=0pt, topsep=2pt]
    \item the b-hadron production cross section
    \item the production fraction of the B meson in question
    \item the integrated luminosity
    \item the two branching fractions listed in Table \ref{tab:decays} for the decay chain in question.
\end{itemize}

\begin{table}
    \centering
    \caption{Estimated production rates for the candidate decay chains}
    \begin{tabular}{ l @{$\to$ } l @{$\to$ } l D{.}{ }{1} D{.}{ }{1} D{.}{.}{5}}
        \noalign{\medskip}\hline\noalign{\smallskip}
        \multicolumn{3}{l}{ } & \multicolumn{3}{c}{production rate (per fb$^{-1}$)*} \\[1pt]
        \multicolumn{3}{c}{decay chain} & \multicolumn{1}{c}{\text{ATLAS}} & \multicolumn{1}{c}{\text{LHCb}} &   \multicolumn{1}{c}{\text{Belle II}} \\
        \noalign{\smallskip}\hline\noalign{\smallskip}
        B$^+$ & $\rho(770)^+\gamma$ & $\pi^+\gamma\gamma$ & 20 & 11 & 0.0003 \\
        B$^+$ & K*(892)$^+\gamma$ & K$^+\gamma\gamma$ & 1800 & 1000 & 0.022 \\
        B$^+$ & K$^*_2(1430)^+\gamma$ & K$^+\gamma\gamma$ & 1500 & 800 & 0.019 \\
        \noalign{\smallskip}\hline\noalign{\smallskip}
        B$^0$ & $\rho(770)^0\gamma$ & $\pi^0\gamma\gamma$ & 19 & 10 & 0.0002 \\
        B$^0$ & $\rho(770)^0\gamma$ & $\eta\gamma\gamma$ & 12 & 6 & 0.0001 \\
        B$^0$ & $\omega(782)\gamma$ & $\pi^0\gamma\gamma$ & 1700 & 900 & 0.020 \\
        B$^0$ & $\omega(782)\gamma$ & $\eta\gamma\gamma$ & 9 & 5 & 0.0001 \\
        B$^0$ & K*(892)$^0\gamma$ & K$^0\gamma\gamma$ & 4700 & 2600 & 0.055 \\
        B$^0$ & K$^*_2(1430)^0\gamma$ & K$^0\gamma\gamma$ & < 500 & < 270 & < 0.006 \\
        \noalign{\smallskip}\hline\noalign{\smallskip}
        B$^0_\mathrm{s}$ & $\phi$(1020)$\gamma$ & $\eta\gamma\gamma$ & 5000 & 2700 & \multicolumn{1}{c}{\text{---}} \\
        B$^0_\mathrm{s}$ & $\phi$(1020)$\gamma$ & $\pi^0\gamma\gamma$ & 510 & 270 & \multicolumn{1}{c}{\text{---}} \\
        \noalign{\smallskip}\hline\noalign{\smallskip}
        \multicolumn{6}{l}{\footnotesize{*Using data in Table \ref{tab:yieldata}; also see text}} \\
    \end{tabular}
    \label{tab:yields}
\end{table}

\noindent Table \ref{tab:yields} gives the results, using the cross sections and production fractions listed in Table \ref{tab:yieldata}. Integrated luminosity cannot be determined in the absence of information on triggers, so the production rates are quoted in units of per fb$^{-1}$ of integrated luminosity. (For reference, the ATLAS detector recorded an integrated luminosity of about 140 fb$^{-1}$ during Run 2.)

Each B meson listed has at least one two-step radiative decay path with an estimated production rate in the detectors at the LHC of over 1000 per fb$^{-1}$. Rates are lower at Belle II, but this is offset by the high luminosity: Belle II aims for an eventual total integrated luminosity of 50~ab$^{-1}$ (5 $\times 10^4$ fb$^{-1}$). Also, the lower background rate and availability of a tag B meson at Belle II are expected to help. On the other hand, B$_{\mathrm{s}}$ mesons are not available to Belle II. Detailed simulation of the processes and detectors is required to indicate the best options.

\begin{table}
    \centering
    \caption{Input data for Table \ref{tab:yields}}
    \begin{tabular}{ l c l l l }
        \noalign{\medskip}\hline\noalign{\smallskip}
         & b-hadron & \multicolumn{3}{c}{B-meson production}\\
         & production & \multicolumn{3}{c}{fraction \citep{Wo22}} \\
         & cross section & \multicolumn{1}{c}{B$^+$} & \multicolumn{1}{c}{B$^0$} & \multicolumn{1}{c}{B$_{\mathrm{s}}$} \\
        \noalign{\smallskip}\hline\noalign{\smallskip}
        ATLAS & 113 $\mu$b*$^{\dag}$ & 0.408 & 0.408 & 0.100 \\
        LHCb & 61 $\mu$b*$^{\ddag}$ & 0.408 & 0.408 & 0.100 \\
        Belle II & 1.11 nb$^{\mathsection}$ & 0.514 & 0.486 & 0  \\
        \noalign{\smallskip}\hline\noalign{\smallskip}
        \multicolumn{5}{l}{\footnotesize{*Using \citep{Ca12} with centre-of-mass energy of 13.0 TeV,}} \\
        \multicolumn{5}{p{6.8cm}}{\footnotesize{$^{\dag}$also transverse momentum $p_\mathrm{T}>$ 5 GeV and magnitude of pseudorapidity $|\eta| <$ 2.5.}} \\
        \multicolumn{5}{l}{\footnotesize{$^{\ddag}$also $
        p_\mathrm{T}>$ 3.5 GeV and 2 $<\eta <$ 5 (e.g. \citep[${\mathsection}2$]{LH17}).}} \\
        \multicolumn{5}{l}{\footnotesize{$^{\mathsection}$From \citep[p. 48]{Ko18}.}} \\
    \end{tabular}
    \label{tab:yieldata}
\end{table}

\subsection{Signals and Backgrounds}
\label{sec:SigBack} 
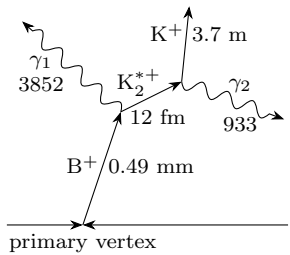
\begin{figure}
    \centering
    \input{fig_signal}
    \caption{Sketch of the two-step radiative decay of a B$^+$ meson (not to scale). Distances are $c\tau$ values or $h/\Gamma$ for the K$^{*+}_2$ ($\tau$: mean life, $\Gamma$: full width); photon energies are in MeV \citep{Wo22}.}
    \label{fig:signal}
\end{figure}

\noindent
Figure \ref{fig:signal} is a sketch of the signal event for the decay of a B$^+$ meson via a K$^{*+}_2$, the lifetime of which is much too short for the vertex separation to be detectable; that is, the two photons appear to come from a common vertex. The first step in the decay sequence, the reconstruction of the B$^+$ mass from the high-energy photon and the K$^{*+}_2$, was demonstrated both by the CLEO and BaBar Collaborations \citep{Co00, Au04} in the course of their branching-fraction measurements. Their method includes techniques for separating decays through the K$^*_2(1430)^+$ from those through the nearby K$^*(1410)^+$, which has a comparable branching fraction, and for handling a range of other likely backgrounds.

The second step in the decay chain involves reconstructing the K$^{*+}_2$ from its decay products. This may prove more difficult, owing to the lower photon energy of 933 MeV. It has not been demonstrated previously: the measurement leading to the K$^{*+}_2$ branching fraction quoted by the Particle Data Group \citep{Wo22} was not of radiative decay but rather of the radiative width in excitation by Primakoff production \citep{Ci82}. (The similar measurement by the KTeV Collaboration leading to the upper limit for the K$^{*0}_2$ radiative-decay branching fraction in \citep{Wo22} is reported in \citep{Al02}, which also contains a useful spectral diagram of the neutral K resonances.)

Again, detailed modelling of the processes in current detectors is indicated: to test the feasibility of reconstructing the K$^{*+}_2$ and of separating signal events from the many backgrounds.

\subsection{Finite-Solid-Angle Corrections}
\label{sec:Omega} \noindent
The schematic diagrams in Figs \ref{fig:angdist} and \ref{fig:angcorr} show arrangements typical of nuclear-physics experiments, using gamma-ray detectors that subtend significant solid angle at the beam spot. This has the effect of attenuating the angular variation; that is, Eq. \ref{angdist} becomes (e.g. \citep{Ro53, Ha66})
\begin{equation} \label{expangdist}
    W(\theta) = A_0[1 + Q_2a_2P_2(\cos\theta) + Q_4a_4P_4(\cos\theta)+\ldots].
\end{equation}
\noindent
During the 1960s and 1970s significant effort was invested in computing and tabulating the attenuation factors $Q_k$ using methods that were precursors of \textsc{geant4} (e.g. \citep{Ha66}). The accuracy of the calculations was tested in precision experiments using decay sequences with spins that had previously been determined (e.g \citep{Br80}).

Values of $Q_k$ depend on photon energy as well as experimental geometry, detector dimensions and detector material. A corresponding correction would be required in the experiment proposed herein. However, since an individual segment of a modern segmented electromagnetic calorimeter subtends a smaller solid angle at the interaction region than was typical in nuclear-physics experiments of the 1960s and 1970s, the attenuation should be less; that is, the $Q_k$ factors may be expected to be closer to unity than for a large unsegmented gamma-ray detector located close to the beam spot.

\subsection{In the Event of an Unexpected Result}
\label{sec:Nonzero} \noindent
It is of course expected that, should the experiment turn out to be feasible, it will find a result in agreement with the SM prediction of spin zero. The contrary would be indicated by one of the following experimental outcomes:
\begin{itemize}
    \item In decays via a spin-1 particle, a nonzero B-meson spin results in an $a_2$ value less than 0.5, as shown in Fig. \ref{fig:spin1}. A sufficiently negative value, less than $-1/5$, would rule out spin two.
    \item In decays via K$^*_2$, an $a_4$ much less than 8/7 indicates nonzero $J_0$. The $a_4$ value is negative for $J_0=1$ and positive for $J_0=2$. This is sufficient to discriminate between these two spins except when $a_4=0$, but that value corresponds to $\delta_1=0$, for which the sign of $a_2$ discriminates between the two spins.
\end{itemize}

\noindent Should a nonzero spin be found, it would be wise to add curves for higher $J_0$ values to Figs \ref{fig:spin1}--\ref{fig:spin24}, to check whether the experimental results are sufficient to rule these out. It may be necessary to combine results from several decay chains to obtain a unique and model-independent $J_0$ value, as was frequently the experience in nuclear spectroscopy.

\subsubsection{Mixing Ratios}
\label{sec:Mixing} \noindent
When $J_0$ is nonzero, the method yields a value for the mixing ratio $\delta_1$ in addition to the spin. In that case, measurements should be made on as many decay chains in Table~\ref{tab:decays} as possible; for mixing ratios depend on the states involved, so are very likely different for different intermediate particles. This should provide valuable guidance for theory, as it does in nuclear physics.

\subsubsection{Parity}
\label{sec:Parity} \noindent
A nonzero B-meson spin would enable a measurement of the parity of that particle. Model-independent measurement of parity has also been standard practice in nuclear spectroscopy for many decades, using a method related to the method for measuring spin. However, the method fails for decays from a spin-zero state, as the following paragraphs explain.

The technique involves the measurement of the polarisation of radiation either from or to the state in question, since this depends on the parity difference between initial and final state. That is, a measurement of parity $P$ uses only the first transition in the cascade of Fig. \ref{fig:cascade}: the configuration of Fig. \ref{fig:angdist} rather than Fig. \ref{fig:angcorr}, effectively. It has been known since the early days of quantum electrodynamics that the differential cross sections of both Compton scattering and pair production have a dependence on the polarisation of the incident photon \citep{Kl29,Be34}. Various experimental arrangements for exploiting this to construct a gamma-ray polarimeter have been published (e.g. \citep{Ko72,Ma81}).

It is usual to define the degree of polarisation $\mathcal{P}$ of a $\gamma$ ray as
\begin{equation} \label{poldef}
    \mathcal{P}(\theta)=\frac{W(\theta,0)-W(\theta,90\degree)}{W(\theta,0)+W(\theta,90\degree)},
\end{equation} \noindent
where the second argument of the intensity $W$ refers to the angle between the electric vector of the photon and the plane defined by the beam direction and the photon's direction of propagation. Radiative decays are classified as either electric (E) or magnetic (M), and these have opposite polarisation. Electric transitions with multipolarity $L$ between states $|i\rangle$ and $|f\rangle$ satisfy the selection rule $P_iP_f=(-)^L$, with the opposite applying to magnetic transitions. Radiative decay theory gives, for a mixed quadrupole--dipole transition (e.g. \citep{Po67})
\begin{equation} \label{polval}
    \mathcal{P}(\theta)=\pm \frac{(a_2-b_2)P_2^{(2)}(\cos\theta)-a_4P_4^{(2)}(\cos\theta)/6}
                                 {2[1+a_2P_2(\cos\theta)+a_4P_4(\cos\theta)]} \qquad
    \begin{aligned}
        +&\text{ for M1/E2} \\[-1pt]
        -&\text{ for E1/M2,}
    \end{aligned}
\end{equation} \noindent
where the $a_k$ are given by Eq. \eqref{ak}, $P_L^{(M)}(\cos\theta)$ are associated Legendre functions and
\begin{equation} \label{b2}
    b_2=\frac{8a_2\delta R_2(12J_\mathrm{i}J_\mathrm{f})}{3[R_2(11J_\mathrm{i}J_\mathrm{f})+
        2\delta R_2(12J_\mathrm{i}J_\mathrm{f})+\delta^2R_2(22J_\mathrm{i}J_\mathrm{f})]}.
\end{equation} \noindent
In this equation, the $R_k$ are given by Eq. \eqref{Rk} and $\delta$ is the mixing ratio of the transition.

Conceptually, the measurement involves obtaining $\delta$ from the angular distribution of the radiation, which enables the evaluation of Eqs \eqref{polval} and \eqref{b2} once a value of the measurement angle $\theta$ is settled upon. A sufficiently accurate measurement of $\mathcal{P}$ at angle $\theta$ then determines whether the radiation is M1/E2 or E1/M2, thereby indicating whether the initial and final states have the same or opposite parity.

The difficulty in applying this to the B mesons lies in their expected spin value of zero. If true, then the numerator of Eq. \eqref{polval} is identically zero, rendering the measurement impossible in principle. Should B-meson spin turn out not to be zero, then the possibility of a parity measurement is revived.

\section{Summary and Conclusion}
\label{sec:Conclusion} \noindent
The B$^+$ meson has a two-step radiative decay cascade that can (in principle) yield a model-independent measurement of its spin, enabled by the fact that the intermediate state in question, K$^*_2(1430)^+$, has spin two.

For the B$^0$ meson, the feasibility of a model-independent measurement depends on the value of the branching fraction of the radiative decay of K$^*_2(1430)^0$; at present only an upper limit is known. However, measurements of decays through a $J = 1$ intermediate particle would nevertheless have value as a test of the SM, even though the expected result would not lead to an unambiguous spin value. The same applies for the B$^0_\mathrm{s}$ meson. Nothing can be concluded concerning the B$^+_\mathrm{c}$ meson, owing to lack of information on its radiative decay modes.

It is suggested that these considerations are sufficiently encouraging to justify the effort required for the next steps: the systematic exploration of background processes and the detailed modelling of the signal and backgrounds in modern detectors. Such modelling is needed to give confidence --- or otherwise --- in the feasibility of the measurement, and to indicate the best options.

The quantum theory of angular momentum (e.g. \citep{Ro57, Za98}) is a foundational pillar of quantum mechanics. It has widespread applicability, so it is unsurprising that an angular-momentum based formalism originally developed for nuclear spectroscopy \citep{Fr53, Ro67} and later found useful in the physics of atoms and molecules (e.g. \citep{Fe93}) should also find application in particle physics.

\section*{Acknowledgement} \noindent
Professor P. Jackson (University of Adelaide) is warmly thanked for guidance, encouragement and critical comments on this paper.

Funding: This research was not funded by a specific grant from any funding agency, whether public, commercial, not-for-profit or of any other type. Infrastructure support was provided by the Australian Government through the Australian Research Council Centre of Excellence for Dark Matter Particle Physics (CDM, CE200100008).

\end{document}

%% file: fig_angdist.tex
\begin{tikzpicture}[outer sep = auto]
  \tikzstyle{every node}=[font=\footnotesize]
  \draw[-Stealth] (-1.5,0) -- (0,0) node[anchor=south] {collision region};
  \draw[Stealth-] (0,0) -- (3,0);
  \coordinate (a) at (2.5, -1.4434);
  \draw[-Stealth, decorate, decoration = {coil, aspect = 0, mirror}] (0,0) -- (a);
  \draw (1.05,0) arc (0:-30:1.05);
  \node at (1.13, -0.3) {$\theta$};
  \coordinate (b) at (2.425, -1.5733);
  \coordinate (c) at (2.858, -1.4635);
  \draw [rotate = 60] (b) rectangle (c);
  \node[anchor=east] at (b) {detector};
\end{tikzpicture}

%% file: fig_angcorr.tex
\begin{tikzpicture}[outer sep = auto]
  \tikzstyle{every node}=[font=\footnotesize]
  \draw[very thick] (0.3,-0.3) -- (-0.3,0.3) node[anchor = east] {target};
  \draw[Stealth-] (0,0) -- (2,0);
  \coordinate (a) at (0, -1.5);
  \draw[-Stealth, decorate, decoration = {coil, aspect = 0, mirror}] (0,0) -- (a);
  \draw (-0.15, -1.8) rectangle +(0.3, 0.3);
  \node[anchor = west] at (0.15, -1.65) {detector 1};
  \coordinate (b) at (-1.785, -0.650);
  \draw[-Stealth, decorate, decoration = {coil, aspect = 0}] (0,0) -- (b);
  \draw (0,-0.89) arc (-90:-160:0.89);
  \node at (-0.7, -0.8) {$\theta$};
  \coordinate (c) at (-1.836, -0.5);
  \coordinate (d) at (-2.0, -0.89);
  \draw [rotate = 20] (c) rectangle (d);
  \node[anchor = north] at (-1.9, -0.85) {detector 2};
\end{tikzpicture}

%% file: fig_cascade.tex
\begin{tikzpicture}[outer sep = auto]
  \tikzstyle{every node}=[font=\footnotesize]
  \draw[very thick] (-0.8, 2.6) node[anchor = east] {$|0\rangle$} --
         (0.8, 2.6) node[anchor = west] {$J_{0}$};
  \draw[-Stealth] (0, 2.6) -- (0, 0);
  \node[anchor = west] at (0, 1.3) {$L_{1}$, $L_{1}+1$, $\ldots$};
  \draw[very thick] (-0.8, 0) node[anchor = east] {$|1\rangle$} -- 
         (0.8, 0) node[anchor = west] {$J_{1}$};
  \draw[-Stealth] (0, 0) -- (0, -2);
  \node[anchor = west] at (0, -1) {$L_{2}$, $L_{2}+1$, $\ldots$};
  \draw[very thick] (-0.8, -2) node[anchor = east] {$|2\rangle$} -- 
         (0.8, -2) node[anchor = west] {$J_{2}$};
\end{tikzpicture}

%% file: fig_signal.tex
\begin{tikzpicture}[outer sep = auto]
 \tikzstyle{every node}=[font=\footnotesize]
  \draw[-Stealth] (-1,-0.5) -- (0,-0.5) node[anchor=north] {primary vertex};
  \draw[Stealth-] (0,-0.5) -- (2.7,-0.5);
  \draw[-Stealth] (0,-0.5) -- (0.5,1);
  \node[anchor=east] at (0.33,0.3) {B$^+$};
  \node[anchor=west] at (0.22,0.275) {0.49 mm};
  \draw[-Stealth, decorate, decoration = {coil, aspect = 0, mirror}] (0.5,1) -- (-0.8,2.2);
  \node[anchor=south] at (-0.55,1.45) {$\gamma_1$};
  \node[anchor=north] at (-0.55,1.58) {3852};
  \draw[-Stealth] (0.5,1) -- (1.3,1.4);
  \node[anchor=west] at (0.33,1.39) {K$^{*+}_2$};
  \node[anchor=west] at (0.5,0.94) {12 fm};
  \draw[-Stealth, decorate, decoration = {coil, aspect = 0, mirror}] (1.3,1.4) -- (2.65,0.9);
  \node[anchor=south] at (2.07,1.14) {$\gamma_2$};
  \node[anchor=north] at (2.07, 1.05) {933};
  \draw[-Stealth] (1.3,1.4) -- (1.4,2.4);
  \node[anchor=east] at (1.45,2.05) {K$^+$};
  \node[anchor=west] at (1.32,2.025) {3.7 m};
\end{tikzpicture}